\documentclass{optica-article}

\journal{opticajournal} 

\articletype{Research Article}

\usepackage{lineno}
\usepackage{bm}
\usepackage{siunitx}

\begin{document}

\title{sCWatter: Open source coupled wave scattering simulation for spectroscopy and microscopy}

\author{Ruijiao Sun,\authormark{1} Rohith Reddy,\authormark{1} and David Mayerich\authormark{1,*}}

\address{\authormark{1}Department of Electrical and Computer Engineering, STIM Lab, 4796 Cullen Blvd, Houston, US, 77004. Fax:713-743-4444; Tel:713-743-6105.}

\email{\authormark{*}mayerich@uh.edu} 


\begin{abstract*} 
Several emerging microscopy imaging methods rely on complex interactions between the incident light and the sample. These include interferometry, spectroscopy, and nonlinear optics. Reconstructing a sample from the measured scattered field relies on fast and accurate optical models. Fast approaches like ray tracing and the Born approximation have limitations that are limited when working with high numerical apertures. This paper presents sCWatter, an open-source tool that utilizes coupled wave theory (CWT) to simulate and visualize the 3D electric field scattered by complex samples. The sample refractive index is specified on a volumetric grid, while the incident field is provided as a 2D image orthogonal to the optical path. We introduce connection equations between layers that significantly reduce the dimensionality of the CW linear system, enabling efficient parallel processing on consumer hardware. Further optimizations using Intel MKL and CUDA significantly accelerate both field simulation and visualization.
\end{abstract*}

\footnote{ The software sCWatter is available here: https://github.com/STIM-Lab/scwatter}
\section{Introduction}

Microscopy is becoming increasingly sophisticated by utilizing coherent light with interferometry,\cite{popescu2006diffraction, reddy2013high, reddy2012optimizing} spectroscopy,\cite{alqaheem2020microscopy,kaminska2020spectroscopic, walsh2012label, lotfollahi2022adaptive, reddy2010accurate} and nonlinear optics.\cite{zhang2020recent,hong2021giant} As these methods advance, computational modeling plays a crucial role in reconstructing the shape and chemical properties of complex samples.\cite{davis2010theory,davis2010theory2,reddy2011modeling, barone2021computational,su2021physics}

Microscopes aim to reconstruct features at or below the diffraction limit using high numerical aperture (NA) objectives. While ray tracing is a powerful technique for modeling particle paths, it does not accurately capture diffraction and interference effects that dominate when feature sizes approach the wavelength. A diverse array of theoretical models have been developed to model light interactions using wave-based approaches. Born and Rytov approximations \cite{born1926quantenmechanik,gubernatis1977born} take the incident field as the driving field at each point in a sample, providing a first-order approximation that only accounts for a single scattering event. A multi-layer Born model (MLB) was proposed to model 3D phase microscopy \cite{chen2020multi} that overcomes some of these limitations by decomposing the sample into multiple slabs. However, the accuracy can decrease for highly scattering samples with larger variations in refractive index, where multiple scattering events become more significant.

Mie theory accurately describes the electromagnetic field scattered by a spherical object through a series expansion that can be truncated to the desired precision. This model has been used to reconstruct the absorbance spectra for spheres in mid-infrared spectroscopy,\cite{van2013recovery,berisha2017bim} but cannot account for multiple scattering \textit{between} sub-regions within an object in more complex samples. 

Coupled wave theory (CWT) uses a Fourier decomposition of the sample to generate a linear system used to solve the scattered field. CWT represents the sample and field as a Fourier expansion and has been used to simulate mid-infrared spectroscopy for both layered, \cite{davis2010theory} single-layered (2D) heterogeneous samples,\cite{davis2010theory2, reddy2011modeling} and multiple-layered (3D) samples.\cite{sun2022characterization} While CWT requires solving a potentially large linear system, this Fourier-based approach makes it convenient to simulate any point in an optical system by applying linear transforms corresponding to optical components such as objectives and filters. In addition, large samples can be composed of small patches in two dimensions.

In this paper, we develop an open-source software \textit{sCWatter} that uses CW theory to simulate and visualize the electric field scattered by a 3D sample. We focus specifically on efficiencies that enable parallel processing such as visualizing points in the field. By leveraging parallel computing and high-performance libraries, we are able to simulate the imaging process for a range of complex samples and imaging systems using inexpensive consumer hardware.

\section{Theoretical Model}
The CW-based model represents the electric field and sample in terms of their spatial frequencies. For the field, each spatial frequency is represented by a plane wave proceeding in the direction:
\begin{equation}
    \mathbf{s}=[s_x, s_y, s_z]^T
\label{eqn:direction_vector}
\end{equation}
with the $z$ component constrained by the complex refractive index $n$ of the material:
\begin{equation}
    s_z(\bar{\mathbf{s}}, n)=\sqrt{n^2 - \bar{\mathbf{s}} \cdot \bar{\mathbf{s}}} =\sqrt{n^2 - s_x^2 - s_y^2} 
\label{eqn:direction_vector_z}
\end{equation}
where $\bar{\mathbf{s}}$ is a single spatial frequency in terms of its $x$ and $y$ components:
\begin{equation*}
    \bar{\mathbf{s}}=[s_x, s_y]^T
\end{equation*}
which are constant at any spatial location. The plane wave direction in three dimensions is then calculated based on the spatially-dependent refractive index:
\begin{equation*}
    \mathbf{s}\left( \bar{\mathbf{s}}, n\right) =
    \begin{bmatrix}
        \bar{\mathbf{s}}\\
        \sqrt{n^2 - \bar{\mathbf{s}}\cdot\bar{\mathbf{s}}}
    \end{bmatrix}
    =
    \begin{bmatrix}
        s_x\\
        s_y\\
        \sqrt{n^2 - \left( s_x^2 + s_y^2\right)}
    \end{bmatrix}
\end{equation*}
\textbf{Note:} When it is clear which $\bar{\mathbf{s}}$ is associated with a particular $s_z$ component, we will exclude the spatial frequency term: $s_z(n) = s_z(\bar{\mathbf{s}}, n)$.

\subsection{Electric Field}
An electric field $\mathbf{E}(\mathbf{r})\in \mathbb{C}^3$ with vacuum wavelength $\lambda$ is composed of plane waves with amplitude $\mathbf{p}(\mathbf{s})\in\mathbb{C}^3$:
\begin{equation}
\begin{aligned}
\mathbf{E}(\mathbf{r})= \int \mathbf {p}(\mathbf{s})\textup{exp}\left[ik(\mathbf{s}\cdot \mathbf{r})\right]d\mathbf{s}
\label{eqn:ElectricFourier}
\end{aligned}
\end{equation}
where $\mathbf{r}=[x,y,z]^T$ is a spatial coordinate, $\mathbf{s}$ is a direction vector for the plane wave, and $k=\frac{2\pi}{\lambda}$ is the wavenumber. Note that $\lambda$ is given in terms of the vacuum wavelength ($n=1$), since changes in refractive index are accounted for in $\mathbf{s}$ through the $s_z$ component.

This is the equivalent of the inverse Fourier transform, where the coefficients $\mathbf{p}(\mathbf{s})\in\mathbb{C}^3$ are the complex amplitudes for each plane wave. The electric field $\mathbf{E}(\mathbf{r})$ can be decomposed into individual plane waves $\mathbf{p}(\mathbf{s})$ using the Fourier transform to obtain their associated amplitudes:
\begin{equation}
\mathbf{p}(\mathbf{s}) = \int \mathbf{E}(\mathbf{r})\textup{exp}\left[-ik\left(\mathbf{s} \cdot \mathbf{r}\right)\right]d \mathbf{r}
\label{eqn:inverseE}
\end{equation}
The magnetic field $\mathbf{H}(\mathbf{r})$ is defined identically using the complex coefficients $\mathbf{b}(\mathbf{s})$:
\begin{equation}
\mathbf{b}(\mathbf{s}) = \int \mathbf{H}(\mathbf{r})\textup{exp}\left[-ik\left(\mathbf{s} \cdot \mathbf{r}\right)\right]d \mathbf{r}
\label{eqn:inverseH}
\end{equation}
where $\mathbf{E}(\mathbf{r})$ and $\mathbf{H}(\mathbf{r})$ are orthogonal: $\mathbf{E} \cdot \mathbf{H} = 0$.

\subsection{Wave Discretization}
The CW-based model calculates the amplitude of $M$ discrete frequencies, each represented by a unique $\bar{\mathbf{s}}_m$. 

Given $U$ discrete frequencies along the $x$ axis and $V$ discrete frequencies along the $y$ axis, the direction vector associated with the integer frequency $u,v$ is given by:
\begin{equation}
\bar{\mathbf{s}}_{u,v}= 
\frac{2\pi}{k}
\begin{bmatrix}
\frac {u}{X}\\
\frac {v}{Y}
\end{bmatrix}
\label{eqn:discrete_s}
\end{equation}
where the constants $X$ and $Y$ specify the size of the sample along the $x$ and $y$ directions using the same units as $\lambda$. The individual components $u$ and $v$ are signed integers:
\begin{equation*}
    \begin{array}{lr}
    u \in \left[ -\left \lfloor \frac{U}{2} \right\rfloor,  \left \lfloor \frac{U - 1}{2} \right\rfloor \right]
    &
    v \in \left[ -\left \lfloor \frac{V}{2} \right\rfloor,  \left \lfloor \frac{V - 1}{2} \right\rfloor \right]
    \end{array}
\end{equation*}
To simplify the notation, we use a one-dimensional index $m$ with the following conversion:
\begin{equation*}
    m = \left(v+\left\lfloor \frac{V}{2} \right\rfloor\right) * U + u + \left\lfloor \frac{U}{2} \right\rfloor
\end{equation*}
\begin{equation*}
    v = \left\lfloor\frac{m}{U} \right\rfloor - \left\lfloor \frac{V}{2} \right\rfloor\\
\end{equation*}
\begin{equation*}
    u = m - \left[\left(v+\left\lfloor \frac{V}{2} \right\rfloor\right) * U\right] - \left\lfloor \frac{U}{2} \right\rfloor
\end{equation*}

Note that the use of a uniform grid here is merely a computational convention, and we can choose to reconstruct the field from any convenient set of spatial frequencies $\bar{\mathbf{s}}_m$ by specifying their $s_x$ and $s_y$ components. 

\subsection{External Fields}
\label{sec:scattered_fields}
\begin{figure}[ht!]\centering
\includegraphics[width=0.75\textwidth]{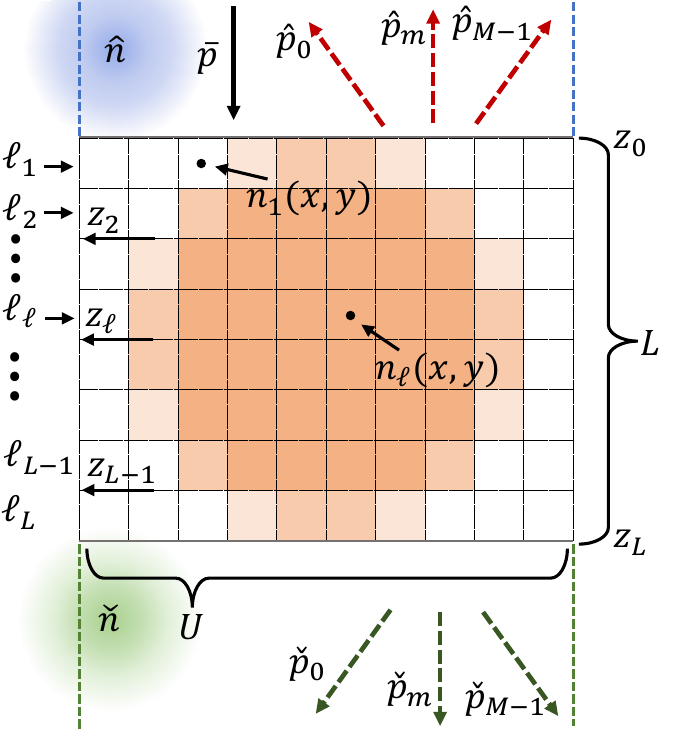}
\caption{The figure shows how a downward incident wave $\bar {\mathbf{P}}$ illuminates on a heterogeneous sample. The reflective waves on the upper boundary are $\hat {\mathbf{P}}$ and the transmitted waves on the lower boundary are $\check {\mathbf{P}}$. The refractive indices of the upper field and lower field are $\hat n$ and $\check n$. The sample is saved as the reciprocal of the sample refractive index $\frac{1}{n^2(\mathbf x_i)}$ at $\mathbf x_i$. The sample contains $L$ layers, and the boundary positions along $z$ direction are represented by $z_{\ell}$ for the $\ell_th$ layer}. 
\label{fig:sample_overview}
\end{figure}

We assume that the sample is bounded above and below by two infinite spaces (Figure \ref{fig:sample_overview}). The top of the sample starts at position $z_0$ and the bottom ends at position $z_L$. The top boundary extends infinitely above the sample and has a real refractive index $\hat n$. The absorbance of this upper layer is assumed to be negligible, and therefore the imaginary part of $\hat{n}$ is zero. The bottom boundary extends infinitely below the sample and has a complex refractive index $\check n$. The sample is discretized into $L$ layers, where each layer has a distribution of refractive indices $n_\ell(x,y)$.

Since most imaging systems are concerned with the field measured outside of the sample, we start by defining three sets of coefficients that determine the external field:
\begin{itemize}
    \item $\bar{\mathbf{p}}(\bar{\mathbf{s}})$ is the amplitude of an incident plane defined at $\bar{z}$ and oriented along direction $\mathbf{s}(\bar{\mathbf{s}}, \hat{n})$.
    
    \item $\hat{\mathbf{p}}(\bar{\mathbf{s}})$ is the amplitude of a plane wave reflected off of the $z_0$ interface at the top of the sample and oriented along direction $\mathbf{s}(\bar{\mathbf{s}}, \hat{n})$.
    
    \item $\check{\mathbf{p}}(\bar{\mathbf{s}})$ is the amplitude of a plane wave transmitted through the sample defined at the $z_L$ interface at the bottom boundary and oriented along direction $\mathbf{s}(\bar{\mathbf{s}}, \check{n})$.
\end{itemize}

Several equations used in this model require scaling plane waves and propagating them between $z$ coordinates, therefore we define the free-space Green's function that propagates a field from $z_a$ to $z_b$:
\begin{equation}
    \begin{aligned}
        G(A, z_a, z_b) = \textup{exp}[ik(z_b - z_a)A]
    \end{aligned}
    \label{eqn:exp}
\end{equation}
with the corresponding tensor function:
\begin{equation}
    \begin{aligned}
        \mathbf{G}(\mathbf{A}, z_a, z_b) = \textup{\textbf{exp}}[ik(z_b - z_a)\mathbf{A}]
    \end{aligned}
    \label{eqn:exp_tensor}
\end{equation}
where $G_{m,n}=\textup{exp}[ik(z_b - z_a)A_{m,n}]$

The plane wave associated with a spatial frequency $\bar{\mathbf{s}}$ at any $z$ position outside of the sample is calculated using the appropriate coefficients:
\begin{equation}
    \mathbf{p}(\bar{\mathbf{s}},z)=
    \begin{cases}
    \begin{array}{lr}
        \bar{\mathbf{p}}(\bar{\mathbf{s}})G\left[s_z(\hat{n}), \bar{z}, z\right] + \hat{\mathbf{p}}(\bar{\mathbf{s}})G\left[s_z(\hat{n}), z_0, z\right] & z<z_0\\
        \check{\mathbf{p}}(\bar{\mathbf{s}})G\left[s_z(\check{n}),z_L, z\right] &z > z_L
    \end{array}
    \end{cases}
\label{eqn:P}
\end{equation}
and integrating to compute the electromagnetic field at any position $(r_x, r_y, z)$:
\begin{equation}
    \mathbf{E}(\mathbf{r}, z) = \sum_{m=1}^{M} \mathbf{p}(\bar{\mathbf{s}}_m, z) \textup{exp}(ik\bar{\mathbf{s}}_m\cdot\bar{\mathbf{r}})
\label{eqn:e_from_p}
\end{equation}
where $\bar{\mathbf{r}} = [r_x, r_y]^T$.

\subsection{Sample discretization}
For an $L$-layer heterogeneous sample (Figure \ref{fig:sample_overview}), each layer is converted to a set of discrete Fourier coefficients:
\begin{equation}
    \phi_{m} ^{(\ell)} = \sum_{\bar{\mathbf{r}}} n^2_{\ell}  (\bar{\mathbf{r}})\textup{exp}\left[-ik \bar{\mathbf{r}}\cdot\bar{\mathbf{s}}_m\right]
\label{eqn: epsilon}
\end{equation}
where $\epsilon_\ell(x,y) = n_\ell^2(x,y)$ is the square of the complex refractive index of the sample at position $x,y$ in layer $\ell$. The Fourier coefficient $\phi_{m} ^{(\ell)}$ corresponds to the spatial frequency $\mathbf{s}_{m}$ in Equation \ref{eqn:discrete_s}. These coefficients are readily obtained for a uniform grid by calculating the squared refractive index at each position and applying the fast Fourier transform.

Our calculations will also consider the reciprocal of the sample refractive index expressed using its spatial Fourier coefficients $\psi_{m} ^{(\ell)}$:

\begin{equation}
    \psi_{m} ^{(\ell)} = \sum_{\bar{\mathbf{r}}} \frac{1}{n^2_{\ell}(\bar{\mathbf{r}})} \textup{exp}\left[-ik \bar{\mathbf{r}}\cdot\bar{\mathbf{s}}_m\right]
\label{eqn:inverse_n_fourier}
\end{equation}

\subsection{Simulation Theory}
Maxwell's equations describe the properties of an electrical field and its relationship to the corresponding magnetic field. Gauss' law:
\begin{equation}
    \triangledown  \cdot \mathbf{E}(\mathbf {r})=\frac{\rho}{\epsilon(\mathbf {r})}
\end{equation}
describes the relationship between the electric field $\mathbf E$, the charge density $\rho$, and the vacuum permittivity $\epsilon(\mathbf {r}, \bar \nu)$. Assuming the sample does not contain an electrical charge, the charge density $\rho$ is zero. Combining this and expressing the field divergence as plane wave components gives us the constraint:
\begin{equation}
    \frac{\partial\mathbf{P}}{\partial x} + \frac{\partial\mathbf{P}}{\partial y} + \frac{\partial\mathbf{P}}{\partial z} = 
    \mathbf{S}^T \mathbf{P} = 0
\label{eqn:gauss_law}
\end{equation}

Faraday's law (Equation \ref{eqn:faraday_law}) represents how a changing magnetizing field $\mathbf H$ can induce an electric field $\mathbf E$. Ampère's law (Equation \ref{eqn:ampere_law}) relates the magnetizing field $\mathbf H$ with the changing electricity field $\mathbf E$
\begin{equation}
\nabla \times \mathbf{E}(\mathbf{r})=i k \sqrt{\frac{\mu_0}{\varepsilon_0}} \mathbf{H}(\mathbf{r}) 
\label{eqn:faraday_law}
\end{equation}
\begin{equation}
\nabla \times \mathbf{H}(\mathbf{r})=-i k n^2(\mathbf{r}) \sqrt{\frac{\varepsilon_0}{\mu_0}} \mathbf{E}(\mathbf{r})
\label{eqn:ampere_law}
\end{equation}

Faraday's Law (Equation \ref{eqn:faraday_law}) and Ampere's Law (Equation \ref{eqn:ampere_law}) require that the electric and magnetic fields at any boundary are continuous. Therefore the difference between the reflected and transmitted electric and magnetic fields must be zero:
\begin{equation}
\mathbf{P}_1 - \mathbf{P}_2 = 0
\label{eqn:boundary_1}
\end{equation}
\begin{equation}
\mathbf{B}_1 - \mathbf{B}_2 = 0
\label{eqn:boundary_2}
\end{equation}

\subsubsection{Define Boundary conditions}
\begin{figure}[ht!]\centering
\includegraphics[width=0.75\textwidth]{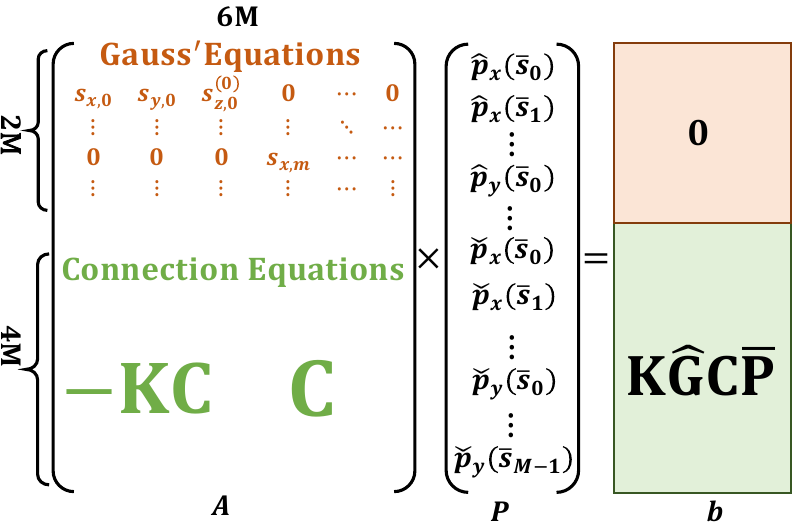}
\caption{The electric field vector $\mathbf P$ at the boundaries is calculated by solving the linear system. For the $6M$ linearly independent conditions, $2M$ linearly independent conditions are provided by Gauss' Equations (Equations \ref{eqn:gauss_law}), and $4M$ by the connection equations mentioned in the next subsection, in which $\mathbf{C}, \mathbf{C}^{\prime} \in \mathbb {C}^{4M\times 3M}$ and $\mathbf{K}, \hat {\mathbf{G}} \in \mathbb {C}^{4M\times4M}$. }
\label{fig:linear_functions_Hete}
\end{figure}

Gauss' Law provides $2M$ constraints (Figure \ref{fig:linear_functions_Hete}) similar to those in the homogeneous example. An additional $4M$ \textit{connection equations} provide the boundary constraints (Equations \ref{eqn:boundary_1} and \ref{eqn:boundary_2}). These equations are calculated by:

\begin{enumerate}
    \item Construct a property matrix $\mathbf D ^{\ell}$ for each layer. 
    \item Calculate the eigenvectors and eigenvalues of $\mathbf{D^{\ell}}$ to express the top and bottom boundary fields for each Fourier coefficient.
    \item Derive the connection equations by enforcing continuity of the boundary field (Equations \ref{eqn:boundary_1} and \ref{eqn:boundary_2}).
    \item Construct and solve the linear system (Figure \ref{fig:linear_functions_Hete}) to derive the external field coefficients $\hat{\mathbf{P}}$ and $\check{\mathbf{P}}$.
\end{enumerate}

\subsubsection{Derive the Property Matrix $\mathbf D$}
Each layer is characterized by an associated property matrix $\mathbf{D}$ that describes how the field changes as it propagates through the layer along the $z$-axis. $\mathbf{D}$ for any layer is derived using the Maxwell-Faraday equation (Equation \ref{eqn:faraday_law}) and Ampère's law(Equation \ref{eqn:ampere_law}), which define the partial derivatives of the electric and magnetic fields along $z$.

Substituting the Equations \ref{eqn:inverseE} and \ref{eqn:inverseH} into the Equations \ref{eqn:faraday_law} and \ref{eqn:ampere_law} produces the following linear system within the layer:
\begin{equation}
\begin{bmatrix}
\frac{\partial \mathbf{P}_x}{\partial z} \\
\frac{\partial \mathbf{P}_y}{\partial z} \\
\frac{\partial \mathbf{B}_x}{\partial z} \\
\frac{\partial \mathbf{B}_y}{\partial z}
\end{bmatrix} = ik\mathbf D\begin{bmatrix}
\mathbf{P}_x\\ 
\mathbf{P}_y\\ 
\mathbf{B}_x\\
\mathbf{B}_y
\end{bmatrix} = ik\mathbf{D}\mathbf{R}
\label{eqn:Phi_system}
\end{equation}
where $\mathbf{R}$ is the vector of transverse component of the electric and magnetic fields. The unknown $M$-dimensional sub-vectors of $\mathbf{R}$ have the format:
\begin{equation}
    \mathbf{P}_x =
    \begin{bmatrix}
         p_x\left(\bar{\mathbf{s}}_0 \right)\\
         p_x\left(\bar{\mathbf{s}}_1 \right)\\
         \vdots\\
         p_x\left(\bar{\mathbf{s}}_{M-1} \right)
    \end{bmatrix}
\end{equation}
The property matrix $\mathbf D$ is a \textbf{known} $4M\times 4M$ matrix computed from the layer's material properties described in Equations \ref{eqn: epsilon} and \ref{eqn:inverse_n_fourier}:
\begin{equation}
    \mathbf{D}=
    \begin{bmatrix}
        \mathbf{0} & \mathbf{0} & \mathbf{s}_x\mathbf{s}_y^T\circ \bm \Psi & \mathbf{I} - \mathbf{s}_x\mathbf{s}_x^T\circ \bm \Psi\\
        \mathbf{0} & \mathbf{0} & \mathbf{s}_y\mathbf{s}_y^T\circ \bm \Psi - \mathbf{I} & - \mathbf{s}_y\mathbf{s}_x^T\circ \bm \Psi\\
        \mathbf{s}_x\mathbf{s}_y^T & \bm \Phi-\mathbf{s}_x\mathbf{s}_x^T & \mathbf{0} & \mathbf{0}\\
        \mathbf{s}_y\mathbf{s}_y^T - \bm \Phi & -\mathbf{s}_y\mathbf{s}_x^T & \mathbf{0} & \mathbf{0}
    \end{bmatrix}
    \label{eqn:matrix_D}
\end{equation}
The vectors $\mathbf s_x, \mathbf s_y \in \mathbb{R}^{M\times 1}$ are the spatial frequencies and $\circ$ is the Hadamard product.

The material properties for the layer are encoded in the matrices $\bm \Phi\in \mathbb C^{M\times M}$ and $\bm \Psi\in \mathbb C^{M\times M}$. The rows of $\bm \Phi$ and $\bm \Psi$ are the dynamic phase-shifted result of $\bm \phi$  (Equation \ref{eqn: epsilon}) and $\bm \psi$  (Equation \ref{eqn:inverse_n_fourier}), which represent convolution with the boundary field $\mathbf P$ using a circulant matrix:
\begin{equation*}
    \bm \Psi =
    \begin{bmatrix}
        \color{blue}{\psi_0}               & \psi_{M-1}           & \cdots               & \psi_{\lfloor{\frac{M}{2}}\rfloor}   & \cdots     & \psi_1\\
        \psi_1               & \color{blue}{\psi_0}             & \cdots               & \psi_{\lfloor{\frac{M}{2}}\rfloor+1} & \cdots     & \psi_2\\
        \vdots               & \vdots               & \ddots               & \vdots               & \ddots     & \vdots\\
        \psi_{\lfloor{\frac{M}{2}}\rfloor-1} & \psi_{\lfloor{\frac{M}{2}}\rfloor-2} & \cdots               & \psi_{M-1}           & \cdots     & \psi_{\lfloor{\frac{M}{2}}\rfloor}\\
        \psi_{\lfloor{\frac{M}{2}}\rfloor}   & \psi_{\lfloor{\frac{M}{2}}\rfloor-1} & \cdots               & \color{blue}{\psi_0}             & \cdots     & \psi_{\lfloor{\frac{M}{2}}\rfloor+1}\\
        \psi_{\lfloor{\frac{M}{2}}\rfloor+1} & \psi_{\lfloor{\frac{M}{2}}\rfloor}   & \cdots               & \psi_{1}             & \cdots     & \psi_{\lfloor{\frac{M}{2}}\rfloor+2}\\
        \vdots               & \vdots               & \ddots               & \vdots               & \ddots     & \vdots\\
        \psi_{M-1}           & \psi_{M-2}           & \cdots               & \psi_{\lfloor{\frac{M}{2}}\rfloor-1} & \cdots     & \color{blue}{\psi_0}
    \end{bmatrix}
\end{equation*}

\subsubsection{Eigendecomposition of $\mathbf D$}
The eigendecomposition for each $\mathbf D$ generates a list of $\Gamma$ and $\mathbf{Q}$, represented as 
$$\mathbf D= \mathbf Q\bm \Gamma\mathbf Q^{-1}$$
where $ \text{diag} \left(\bm\Gamma\right) = \left[\gamma_1, -\gamma_1, \gamma_2, -\gamma_2, \cdots, \gamma_{2M}, -\gamma_{2M}\right]^T$ and all $\gamma _i\geq 0$ and $\gamma_i \geq \gamma_{i+1}$. Positive eigenvalues correspond to the field propagating along the positive $z$-axis, while negative eigenvalues correspond to the reflected field within a layer.
%

We use these terms to express the linear system in Equation \ref{eqn:Phi_system} as a set of linear equations:
\begin{equation}
    \frac{\partial\mathbf{R}}{\partial z} = ik\mathbf{Q}\bm\Gamma\mathbf{Q}^{-1}\mathbf{R}
\end{equation}
Solving this partial differential equation for $\mathbf{R}$ yields:
\begin{equation*}
    \mathbf{Q}^{-1}\frac{\partial\mathbf{R}}{\partial z} = ik\bm\Gamma\mathbf{Q}^{-1}\mathbf{R}
\end{equation*}
\begin{equation*}
    \frac{\partial\left(\mathbf{Q}^{-1}\mathbf{R}\right)}{\partial z} = ik\bm\Gamma\left(\mathbf{Q}^{-1}\mathbf{R}\right)
\end{equation*}
Since the vector differential equation $\frac{\partial\mathbf{a}}{\partial z} = \mathbf{a}$ has the solution $a_j=\alpha_je^z$, the resulting vector is composed of the scalar components:
\begin{equation*}
    \left[\mathbf{Q}^{-1}\mathbf{R}\right]_j=\alpha_j  \exp\left[ ik\Gamma_{jj}z \right]
\end{equation*}
Multiplying by $\mathbf{Q}$ provides the electromagnetic field:
\begin{equation}
   \mathbf{R} =  \mathbf{Q}\left[\text{\bf exp} \left( ik\bm\Gamma z\right)\right]\bm\alpha 
\label{eqn:field_in_layer}
\end{equation}
where {\bf exp} is an element-wise exponential function and $\bm \alpha \in \mathbb{C}^{4M}$ is a vector of unknown quantities resulting from the integration.

We then break $\mathbf Q$ into two matrices with columns corresponding to positive ($\check {\mathbf Q} \in \mathbb{C}^{4M\times 4M}$) and negative ($\hat {\mathbf Q} \in \mathbb{C}^{4M\times 4M}$) eigenvalues:
\begin{align*}
    \mathbf{Q} &= {\mathbf Q} \mathbf{I} = {\mathbf Q} \left( \check{\mathbf{I}} + \hat{\mathbf{I}} \right) = \check {\mathbf Q} + \hat {\mathbf Q}
\end{align*}
where $\check{\mathbf I}[2j,2j] = 1$ and $\hat{\mathbf I}[2j+1,2j+1] = 1$ for $j\in [0,2M-1]$.

\subsubsection{Derive Connection Equations }
\begin{figure}[ht!]\centering
\includegraphics[width=0.75\textwidth]{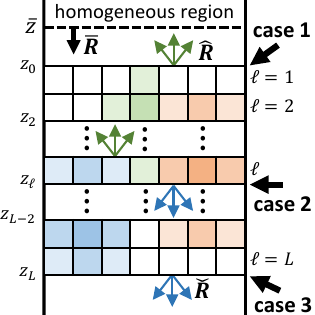}
\caption{The continuity at the boundaries has three cases. Case $1$ represents the boundary conditions between the upper homogeneous field and the first sample layer; Case $2$ shows the continuity between two internal adjacent sample layers; Case $3$ represents the boundary conditions between the last sample layer and the lower homogeneous layer.}
\label{fig:HeteBoundary}
\end{figure}

For each layer, Equation \ref{eqn:field_in_layer} relates the transverse components of the electric and magnetic fields to the eigendecomposition of that layer's property matrix $\mathbf{D}^{(\ell)}$ and an unknown vector $\bm\alpha^{(\ell)}$. In this section, we formulate a set of connection equations that remove the unknown $\bm {\alpha}$ from these equations. These connection equations are derived from three cases (Figure \ref{fig:HeteBoundary}):

\begin{itemize}
    \item The \textbf{upper boundary} constraint accounts for the transition from a homogeneous region into the heterogeneous sample (Figure \ref{fig:HeteBoundary}a). These constraints are derived from Equations \ref{eqn:P} and \ref{eqn:field_in_layer} to produce:
     \begin{equation*}
    \bar{\mathbf{G}}\bar{\mathbf{R}} + \hat{\mathbf{R}}=\mathbf{\check Q}^{(1)}\bm \alpha^{(1)}+\mathbf{\hat Q}^{(1)}\mathbf{G}\left(\bm \Gamma^{(1)}, z_1, z_0\right)\bm\alpha^{(1)} \end{equation*}
    
    \begin{equation}
        \bar{\mathbf{G}}\bar{\mathbf{R}} + \hat{\mathbf{R}}=\left[\check{\mathbf{Q}}^{(1)} + \hat{\mathbf{Q}}^{(1)}\mathbf{G}\left(\bm \Gamma ^{(1)}, z_1, z_0\right)\right]\bm \alpha^{(1)}
    \label{eqn:upper_boundary}
    \end{equation}
    where the left term is derived from Equation \ref{eqn:P} and the right is derived from Equation \ref{eqn:field_in_layer} when $z=z_0$. The sub-vector $\bar{\mathbf{R}} \in \mathbb{C}^{4M}$ consists of the known components of $\mathbf{R}$ given as as the incident field, and $\hat{\mathbf{R}}\in \mathbb{C}^{4M}$ consists of the unknown components of $\mathbf{R}$ reflected by the sample. 
    $\bar{\mathbf{G}}\in\mathbb{C}^{4M\times4M}$ is a diagonal matrix that propagates the incident field $\bar{\mathbf{R}}$ to the top layer of the sample: 
    $$
    \bar{\mathbf{G}} = \mathbf{G}(\mathbf{s}_z, \bar{z}, z_0)
    $$
    and $\mathbf{s}_z\in\mathbb{R}^M$ is a vector of $s_z$ components for each spatial frequency.


    \item The \textbf{middle boundary} constraints account for the transition from sample layers $\ell-1$ to $\ell$ (red circle in the Figure \ref{fig:HeteBoundary}b). This produces $4M$ equations derived from \ref{eqn:P} and \ref{eqn:field_in_layer}. The $M$ constraints obtained from Equation \ref{eqn:field_in_layer} have the following form:
    
    \begin{equation}
        \begin{aligned}
        &\left[\mathbf{\check Q}^{(\ell-1)}\mathbf{G}\left(\bm \Gamma^{(\ell-1)}, z_{\ell-2}, z_{\ell-1}\right)+\mathbf{\hat Q}^{(\ell-1)}\right]\bm\alpha^{(\ell-1)}  \\=&\left[\mathbf{\check Q}^{(\ell)}+\mathbf{\hat Q}^{(\ell)}\mathbf{G}\left(\bm \Gamma^{(\ell)}, z_{\ell}, z_{\ell-1}\right)\right]\bm\alpha^{(\ell)}  
        \end{aligned}
    \label{eqn:middle_boundary}
    \end{equation}
    
    
    \item The \textbf{lower boundary} constraint accounts for the transition from the last heterogeneous sample layer to the lower homogeneous region (Figure \ref{fig:HeteBoundary}c). This produces $4M$ equations derived from Equations \ref{eqn:P} and \ref{eqn:field_in_layer}:
    \begin{equation}
        \check{\mathbf{R}}=\left[\mathbf{\check Q}^{(L)}\mathbf{G}\left(\bm \Gamma^{(L)}, z_{L-1}, z_{L}\right)+\mathbf{\hat Q}^{(L)}\right]\bm\alpha^{(L)}
        \label{eqn:lower_boundary}
    \end{equation}
    %
    
\end{itemize}

The Connection Equations are created by repeatedly substituting Equations \ref{eqn:upper_boundary} and \ref{eqn:middle_boundary} into Equation \ref{eqn:lower_boundary} to eliminate $\bm \alpha$. We define the following terms to simplify the expansion:
\begin{equation*}
    \Grave{\mathbf{Q}}_{\ell}=\mathbf{\check Q}^{(\ell)} + \mathbf{\hat Q}^{(\ell)}\mathbf{G}\left(\bm \Gamma^{(\ell)}, z_{\ell}, z_{\ell-1}\right)
\end{equation*}
\begin{equation*}
    \Acute{\mathbf{Q}}_{\ell}=\mathbf{\check Q}^{(\ell)}\mathbf{G}\left(\bm \Gamma^{(\ell)}, z_{\ell-1}, z_{\ell}\right) + \mathbf{\hat Q}^{(\ell)}
\end{equation*}
Equations \ref{eqn:upper_boundary} and \ref{eqn:middle_boundary} are re-written:
\begin{align*}
    \bar{\mathbf{G}}\bar{\mathbf{R}} + \hat{\mathbf{R}}= \Acute{\mathbf{Q}}_1\bm \alpha^{(1)} &\rightarrow \bm \alpha^{(1)} = \Acute{\mathbf{Q}}_1^{-1} \left[\bar{\mathbf{G}}\bar{\mathbf{R}} + \hat{\mathbf{R}} \right]\\
    \Grave{\mathbf{Q}}_{\ell-1}\bm\alpha^{(\ell-1)}  =\Acute{\mathbf{Q}}_{\ell}\bm\alpha^{(\ell)} &\rightarrow \bm\alpha^{(\ell)} = \Acute{\mathbf{Q}}_{\ell}^{-1}\Grave{\mathbf{Q}}_{\ell-1}\bm\alpha^{(\ell-1)}
\end{align*}
Finally, these equations are substituted into Equation \ref{eqn:lower_boundary} to create the expansion:
\begin{equation}
 \begin{aligned}
    \check{\mathbf{R}}&=\Grave{\mathbf{Q}}_L\bm\alpha^{(L)}\\
    \check{\mathbf{R}}&=\Grave{\mathbf{Q}}_L\Acute{\mathbf{Q}}_{L}^{-1}\Grave{\mathbf{Q}}_{L-1}\bm\alpha^{(L-1)}\\
    \check{\mathbf{R}}&=\Grave{\mathbf{Q}}_L\Acute{\mathbf{Q}}_{L}^{-1}\Grave{\mathbf{Q}}_{L-1}\cdots \Grave{\mathbf{Q}}_2\Acute{\mathbf{Q}}_{2}^{-1}\Grave{\mathbf{Q}}_{1}\bm\alpha^{(1)}\\
    \check{\mathbf{R}}&=\Grave{\mathbf{Q}}_L\Acute{\mathbf{Q}}_{L}^{-1}\Grave{\mathbf{Q}}_{L-1}\cdots \Grave{\mathbf{Q}}_2\Acute{\mathbf{Q}}_{2}^{-1}\Grave{\mathbf{Q}}_{1}\Acute{\mathbf{Q}}_1^{-1} \left[\bar{\mathbf{G}}\bar{\mathbf{R}} + \hat{\mathbf{R}} \right]
    \label{eqn:connection_R}
\end{aligned}   
\end{equation}

This linear system provides $4M$ constraints on the transverse components of the electromagnetic field $\mathbf{R}=\left[\mathbf{P}_x, \mathbf{P}_y, \mathbf{B}_x, \mathbf{B}_y\right]^T$. We are interested in the electric field $\mathbf{P}=\left[\mathbf{P}_x, \mathbf{P}_y, \mathbf{P}_z\right]^T$, which can be transformed into $\mathbf{R}$ using the matrix $\mathbf{C}\in\mathbb{C}^{4M\times3M}$:
\begin{equation}
    \mathbf{C}\mathbf{P} = \begin{bmatrix}
        \mathbf{I} & \mathbf{0} & \mathbf{0}\\
        \mathbf{0} & \mathbf{I} & \mathbf{0}\\
        \mathbf{0} & \text{diag}\left[-\hat{\mathbf{s}}_z \right] & \text{diag}\left[-\mathbf{s}_y\right]\\
        \text{diag}\left[\hat{\mathbf{s}}_z \right] & \mathbf{0} & -\mathbf{s}_x
    \end{bmatrix}
    \begin{bmatrix}
        \mathbf{P}_x\\
        \mathbf{P}_y\\
        \mathbf{P}_z
    \end{bmatrix}
    =   \mathbf{R}
\end{equation}


Substituting the $\mathbf{R}$s in Equation \ref{eqn:connection_R} with $\mathbf{C} \mathbf P$, we can get the final Connection Equations:
\begin{equation}
    \begin{aligned}
        \mathbf{C}\check{\mathbf{P}}&=\Grave{\mathbf{Q}}_L\Acute{\mathbf{Q}}_{L}^{-1}\Grave{\mathbf{Q}}_{L-1}\cdots \Grave{\mathbf{Q}}_2\Acute{\mathbf{Q}}_{2}^{-1}\Grave{\mathbf{Q}}_{1}\Acute{\mathbf{Q}}_1^{-1} \left[\bar{\mathbf{G}}\mathbf{C}\bar{\mathbf{P}} + \mathbf{C}\hat{\mathbf{P}} \right]\\
        \mathbf{C}\check{\mathbf{P}}&=\mathbf{K} \left[\bar{\mathbf{G}}\mathbf{C}\bar{\mathbf{P}} + \mathbf{C}\hat{\mathbf{P}} \right]\\
        \mathbf{C}\check{\mathbf{P}} - \mathbf{K}\mathbf{C}\hat{\mathbf{P}}&=\mathbf{K}\bar{\mathbf{G}}\mathbf{C}\bar{\mathbf{P}}\\
    \end{aligned}
    \label{eqn:connection_equations}
\end{equation}
 where $\check {\mathbf P}, \bar {\mathbf P}$ and $ \hat {\mathbf P} \in \mathbb{C}^{3M \times 1}$, $\grave {\mathbf Q}_{L}, \Acute {\mathbf Q}_{L}^{-1}, \hdots, \Acute {\mathbf Q}_{1}^{-1}=\mathbf{K} \in \mathbb{C}^{4M \times 4M}$. 
 
 This system of linear equations resolves the final set of constraints used to solve for the external fields $\hat{\mathbf{P}}$ and $\check{\mathbf{P}}$ (Figure \ref {fig:linear_functions_Hete}).

\subsection{Calculate the Scattered Field}

\subsubsection{Outside of the Sample}
The electric field outside of the sample is calculated directly from Equations \ref{eqn:P} and \ref{eqn:e_from_p} for any spatial coordinate $(r_x, r_y, z)$, where $z$ is outside of the sample ($z \leq z_0$ or $z_L \leq z$).

\subsubsection{Inside the Sample}
The interior sample field depends on the $\bm{\alpha}$ values associated with the corresponding layer. If $z_{\ell-1} \leq z \leq z_{\ell}$, then the coordinate is inside layer $\ell$ and requires $\bm{\alpha}^{(\ell)}$. This can be directly calculated from Equation \ref{eqn:middle_boundary} by solving the linear system based on $\bm \alpha$ values from two adjacent layers. Therefore, calculating the internal field at an arbitrary point requires solving all $L$ $\bm\alpha$ vectors as a pre-processing step. The associated plane waves within the layer are calculated from Equation \ref{eqn:field_in_layer} and \ref{eqn:P}:
\begin{equation}
    \mathbf{R}^{(\ell)}(z) =  \left[\mathbf{\check Q}^{(\ell)}\mathbf{G}\left(\bm {\Gamma}^{(\ell)}, z_{\ell-1}, z\right) +\mathbf{\hat Q}^{(\ell)}\mathbf{G}\left(\bm {\Gamma}^{(\ell)}, z_{\ell}, z\right)\right]\bm \alpha^{(\ell)}
    \label{eqn:InsideField}
\end{equation}
The two propagation matrices are the only terms dependent on the spatial coordinate, so to avoid confusion with their coordinates we will redefine them:
\begin{align*}
    \check{\mathbf{G}}^{(\ell)}(z) &= \mathbf{G}\left(\bm {\Gamma}^{(\ell)}, z_{\ell-1}, z\right)\\
    \hat{\mathbf{G}}^{(\ell)}(z) &= \mathbf{G}\left(\bm {\Gamma}^{(\ell)}, z_{\ell}, z\right)
\end{align*}
The plane waves associated with the field at any $z$ coordinate within the sample requires the terms associated with the corresponding layer $\ell$, where $z_{\ell-1} \leq z \leq z_{\ell}$. Within a single layer $\ell$, Equation \ref{eqn:InsideField} is:
\begin{equation}
    \mathbf{R}(z) =  \left[\mathbf{\check Q}\check{\mathbf{G}}(z) +\mathbf{\hat Q}\hat{\mathbf{G}}(z)\right] \bm \alpha
    \label{eqn:layer_R_vector}
\end{equation}

The $x$ and $y$ components of $\mathbf{E}$ are given by $\mathbf{P}_x$ and $\mathbf{P}_y$ directly in $\mathbf{R}$. The $z$ component is calculated from the Fourier coefficients of the magnetic field $\mathbf{B}_x$ and $\mathbf{B}_y$ also stored in $\mathbf{R}$. 

Using Ampere's law (Equation \ref{eqn:ampere_law}) we express the $z$ component of the electric field in terms of the magnetic field:
\begin{equation}
 \left[\mathbf{E}(\mathbf{r})\right]_z=-\frac{1}{i k n^2(\mathbf{r})} \sqrt{\frac{\mu_0}{\varepsilon_0}}\left[\nabla \times \mathbf{H}(\mathbf{r})\right]_z
\label{eqn:electric_z}
\end{equation}
Note that the $z$ component of the curl operator depends only on the $x$ and $y$ components of the magnetic field:
\begin{equation*}
    \left[\nabla \times \mathbf{H}(\mathbf{r})\right]_z = \frac{\partial}{\partial y}H_x - \frac{\partial}{\partial x}H_y
\end{equation*}
Since we have the $x$ and $y$ components of the magnetic field expressed as a Fourier series:
\begin{equation*}
    \left[\mathbf{H}(\mathbf{r})\right]_{xy} =\sqrt{\frac{\varepsilon_0}{\mu_0}} \sum_{m=0}^{M-1} 
    \left[\mathbf{B}_m\right]_{xy}
    \exp \left(ik
    \bar{\mathbf{s}}_m \cdot     \bar{\mathbf{r}}
    \right)
\end{equation*}
calculating the derivatives yeilds:
\begin{equation*}
    \begin{bmatrix}
        \frac{\partial}{\partial x} H_y(\bar{\mathbf{r}})\\
        \frac{\partial}{\partial y} H_x(\bar{\mathbf{r}})
    \end{bmatrix}
    = ik\sqrt{\frac{\varepsilon_0}{\mu_0}} \sum_{m=0}^{M-1}
    \bar{\mathbf{s}}_m \circ 
    \begin{bmatrix}
        b_{y}\\
        b_{x}
    \end{bmatrix}_m
    \exp \left(ik
    \bar{\mathbf{s}}_m \cdot \bar{\mathbf{r}}
    \right)
\end{equation*}
Using these terms in Equation \ref{eqn:electric_z} provides the $z$ component of the electric field in known terms in $\mathbf{R}$ (Equation \ref{eqn:InsideField}):
\begin{equation*}
    E_z(\bar{\mathbf{r}})=\frac{-1}{i k n^2(\bar{\mathbf{r}})} \sqrt{\frac{\mu_0}{\varepsilon_0}}\left[-\frac{\partial}{\partial x}H_y(\bar{\mathbf{r}}) + \frac{\partial}{\partial y}H_x(\bar{\mathbf{r}})\right]
\end{equation*}
Substituting in the plane wave decomposition for the derivatives allows us to eliminate the common constants:
\begin{equation*}
    E_z(\bar{\mathbf{r}})= 
    \left[ \frac{-1}{n^2(\bar{\mathbf{r}})} \right]
    \left[ \sum_{m=0}^{M-1}
    \bar{\mathbf{s}}_m \circ 
    \begin{bmatrix}
        b_{y}\\
        b_{x}
    \end{bmatrix}_m
    \exp \left(ik
    \bar{\mathbf{s}}_m \cdot \bar{\mathbf{r}}
    \right) \right]^T
    \begin{bmatrix}
        -1\\
        1
    \end{bmatrix}
\end{equation*}
Replacing the inverse squared refractive index of the sample with its know Fourier decomposition (Equation \ref{eqn:inverse_n_fourier}):
\begin{multline*}
    E_z(\bar{\mathbf{r}})=
     \left[ -\sum_{m=0}^{M-1}\psi_m \exp \left(ik
    \bar{\mathbf{s}}_m \cdot \bar{\mathbf{r}}
    \right)  \right] \times \\
     \left[ \sum_{m=0}^{M-1}
    \bar{\mathbf{s}}_m \circ 
    \begin{bmatrix}
        b_y\\
        b_x
    \end{bmatrix}_m
    \exp \left(ik
    \bar{\mathbf{s}}_m \cdot \bar{\mathbf{r}}
    \right) \right]^T
    \begin{bmatrix}
        -1\\
        1
    \end{bmatrix}
\end{multline*}
shows that $E_z$ is the product of two Fourier transforms:
\begin{equation*}
    E_z(\bar{\mathbf{r}}) =\mathscr{F}^{-1}\left[ -\bm{\psi} \right]\times
    \left[ \sum_{m=0}^{M-1}
    \bar{\mathbf{s}}_m \circ 
    \begin{bmatrix}
        b_y\\
        b_x
    \end{bmatrix}_m
    \exp \left(ik
    \bar{\mathbf{s}}_m \cdot \bar{\mathbf{r}}
    \right) \right]^T
    \begin{bmatrix}
        -1\\
        1
    \end{bmatrix}
\end{equation*}
By the convolution theorem, the $z$ component of a plane wave oriented along $\mathbf{s}_m$ is:
\begin{equation*}
    \left[p_z\right](\bar{\mathbf{s}}_m) 
        = -\bm{\psi}_m \ast \left\{\bar{\mathbf{s}}_m \cdot
\begin{bmatrix}
    b_{y}\\
    -b_{x}
\end{bmatrix}_m\right\}
    \label{eqn:convolution}
\end{equation*}
which can be represented as a series of nested summations:
\begin{equation}
    \left[P_z\right]_m 
        = -\sum_{m^{\prime}=0}^{M-1}\bm{\psi}_{m^{\prime}} \left\{\left[s_y\right]_{m-m^{\prime}}\left[b_x\right]_{m-m^{\prime}} - \left[s_x\right]_{m-m^{\prime}}\left[b_y\right]_{m-m^{\prime}}\right\}
    \label{eqn:pz_convolution}
\end{equation}

To calculate the electric field at a $z$ coordinate within layer $\ell$, the $\mathbf{R}^{(\ell)}$ vector is computed for the specified $z$ coordinate. The $p_x$ and $p_y$ components are directly extracted for each plane wave, while the $p_z$ component is calculated from the corresponding $b_x$ and $b_y$ values as described previously. Finally, the electric field is recomposed using Equation \ref{eqn:e_from_p}.

\section{Implementation} 
\label{sec:complexity}
We use the following values to specify simulation parameters:
\begin{itemize}
    \item $U \times V \in \mathbb{Z}$ is the grid size used to represent each $z$-axis slice, where the total number of samples is $UV=M$.
    \item $L \in \mathbb{Z}$ is the number of layers along the $z$-axis.
    \item $F \in \mathbb{Z}$ is the number of sample points along $x, y, z$ axes where the field is evaluated to produce the desired model output. $F_{xy} \in \mathbb{Z}$ is the number of sample points for each $x-y$ plane and $F_z \in \mathbb{Z}$ is the number of sample points along $z$ axes. $$F = F_{xy}\times F_z$$
    
\end{itemize}

\subsection{Calculate Fourier Coefficients}
The user specifies the distribution of the sample $\epsilon_{(\ell)}(x,y) \in \mathbb{C}$ in terms of its relative permittivity (square of the complex refractive index). The $U\times V$ coefficients are calculated using Fourier transform. This requires $O\left(LF_{xy}\log_2 \left(F_{xy}\right) \right)$ time using the Fast Fourier transform.

\subsection{Assemble $\mathbf D$ and Calculate Eigendecompositions}
The matrix $\mathbf D^{(\ell)}$ is assembled for each layer (Equation \ref{eqn:matrix_D}). Since $\mathbf D^{(\ell)}\in \mathbb{C}^{4M \times 4M}$, this has a time complexity of $O\left(L M^2\right)$. Since  $\mathbf D^{(\ell)}$ is a complex inverse block diagonal matrix, the resulting eigendecomposition $\mathbf D^{(\ell)} = \mathbf Q^{(\ell)} \bm \Gamma^{(\ell)} \mathbf Q^{(\ell)-1}$ is $O\left(M^3\right)$ using standard algorithms for complex non-symmetric matrices (ex. geev). The assembly and eigendecomposition for the entire sample is $O\left( L M^3\right)$.

\subsection{Build connection equations}
Calculating the expansion in Equation \ref{eqn:connection_R} requires two matrix multiplications for each layer, which has a complexity of $O(LM^2)$.

Then the field vector $\left[\mathbf{P}_x, \mathbf{P}_y, \mathbf{B}_x, \mathbf{B}_y\right]$ needs to be transferred as the unknown electrical field vector $\left[\mathbf{P}_x, \mathbf{P}_y, \mathbf{P}_z\right]$ (Equation \ref{eqn:connection_equations}, which requires $O(M^2)$ time. Therefore the total time for building connection equations is $O({LM^3})$.

\subsection{Solve for the linear system}
According to Figure \ref{fig:linear_functions_Hete}, the first $6L-10$ rows which only need to be filled by linear equations have a time complexity of $O(M)$. The last $4L$ conditions need to be performed in $O(LM)$ time resulting from the $\alpha$ connections for $L$ layers. Assemble the linear system $\mathbf{A}\mathbf{p}=\mathbf{b}$, where $\mathbf{A} \in \mathbb{C}^{6M \times 6M}$, using a linear solver, this requires $O(M^3)$ time. The total time for this step is $O(M^3)$.

\subsection{Calculate the external field}
\label{sec:evaluate_external}
The electric field outside of the sample requires solving Equation \ref{eqn:e_from_p} using $\bar{\mathbf{P}}$, $\hat{\mathbf{P}}$, and $\check{\mathbf{P}}$ from Equation \ref{eqn:P}, which has a time complexity of $O(MF)$.


\subsection{Calculate the internal field}
\label{sec:evaluate_internal}
The vector $\bm{\alpha}^{(\ell)}$ of unknowns arising from integration of the property matrix $\mathbf{D}^{(\ell)}$ must be solved using Equations \ref{eqn:upper_boundary} and \ref{eqn:middle_boundary}, requiring $O(M^3L)$ time. For each $z$ coordinate, the corresponding $\mathbf{R}^{(\ell)}$ vector is calculated (Equation \ref{eqn:layer_R_vector}) in $O(M^3F_z)$ time, where $F_z \leq F$ is the number of distinct $z$ values in the sample points. The $z$ component for each plane wave is calculated by convolving with the sample (Equation \ref{eqn:pz_convolution}), requiring $O(M^2)$ time. Finally, the field at each sample point is calculated using Equation \ref{eqn:e_from_p}, which has a time complexity of $O(MF)$. This requires a total time complexity of $O(M^3F_z + MF)$. However, we discuss in Section \ref{sec:results_field_evaluation} that this can be optimized to obtain $O(M^3L + M^2 F_z + MF)$.

\section{Results}
We first optimized the model and the field visualization by introducing high-performance libraries and CUDA. We then use the model to reveal the accuracy-efficiency trade-off, visualize the field around samples, and collect light-intensity data for more complex samples.

\subsection{Field Evaluation and CUDA Parallelism}
\label{sec:results_field_evaluation}
The most time-consuming step is evaluating the field for a large number of sample points. When modeling the output of an imaging system, only the external field is required (Section \ref{sec:evaluate_external}), which requires $O(MF)$ time to sum all $M$ plane waves at each point.

When the field inside the sample is required (ex. simulating nonlinear optical phenomena), the unknown vector $\bm{\alpha}^{(\ell)}$ must first be solved for each layer (Section \ref{sec:evaluate_internal}), requiring $O(M^3L)$ time. In addition, the $x$ and $y$ components of the electromagnetic field $\mathbf{R}(z)$ must be computed using Equation \ref{eqn:layer_R_vector} for each unique $z$ value being sampled. Finally, the $z$ component of the electric field is calculated by convolving the curl of the magnetic field with the sample (Equation \ref{eqn:pz_convolution}). Directly solving Equation \ref{eqn:layer_R_vector} requires at most $O(M^3F_z)$ time if all the sample points belong to unique layers. However, most of this calculation can be precomputed for each layer by taking advantage of the fact that $\mathbf{G}$ is a diagonal matrix:
\begin{align*}
    \mathbf{R}^{(\ell)}(z) &=  \left[\mathbf{\check Q}^{(\ell)}\check{\mathbf{G}}(z) +\mathbf{\hat Q}^{(\ell)}\hat{\mathbf{G}}(z)\right] \bm \alpha^{(\ell)}\\
     \mathbf{R}^{(\ell)}(z) &= \mathbf{\check Q}^{(\ell)}\check{\mathbf{G}}\bm \alpha^{(\ell)} + \mathbf{\hat Q}^{(\ell)}\hat{\mathbf{G}}\bm \alpha^{(\ell)}\\
    \mathbf{R}^{(\ell)}(z) &= \mathbf{\check Q}^{(\ell)}\left[\text{diag}\left(\bm \alpha^{(\ell)}\right)\right]\check{\mathbf{g}}(z) + \mathbf{\hat Q}^{(\ell)}\left[\text{diag}\left(\bm \alpha^{(\ell)}\right)\right]\hat{\mathbf{g}}(z)
\end{align*}
where $\check g_j = \check G_{jj}$ and $\hat g_j = \hat G_{jj}$ are vectors containing the diagonal nonzero elements of the propagation matrix. This constrains the number of $O(M^3L)$ matrix operations for $L$ layers. The convolution in Equation \ref{eqn:pz_convolution} also takes $O(M^3L)$ time to precalculate $\mathbf{P}_z^{(\ell)}$. Then the internal field $E$ in a single layer $\ell$ can be then calculated combined  Equation \ref{eqn:e_from_p}
\begin{equation}
\begin{aligned}
    &\mathbf{E}^{(\ell)}(\bar{\mathbf{s}},z)= \sum_{m=0}^{M-1} \mathbf{P}^{(\ell)}(m)\\
    =&\sum_{m=0}^{M}\sum_{j=0}^{2M-1}\check{\mathbf{P}}^{(\ell)}_{m, 2j}G(\Gamma_{2j}, z_{\ell-1}, z) + \hat{\mathbf{P}}^{(\ell)}_{m, 2j+1}G(\Gamma_{2j+1}, z_{\ell}, z)
    \end{aligned}
\label{eqn:e_from_p_sample}
\end{equation}

This evaluation, which has a complexity of $O(MF)$ for points outside of the sample (Equation \ref{eqn:e_from_p})  and $O(M^2F)$ for interior points (Equation \ref{eqn:e_from_p_sample}), are the most time-consuming steps since most applications require a large number of sample points. However, this is readily parallelized using CUDA and scaled with the number of CUDA cores, where each kernel calculates the field at each point independently. Due to the parallelism property, \textit{sCWatter} visualize planes instead of volumes, in which case the computational complexity can be reduced to $O(MF_{xy})$ for a $x-y$ plane outside of the sample and $O(M^2F_{xy})$ for an interior $x-y$ plane. The same goes for $x-z$ and $y-z$ planes. Table \ref{tbl:internal_field} shows the evaluation results for an internal plane on a single-threaded CPU (Intel(R) Core(TM) i9-10900F CPU @ 2.80kGHz) and CUDA kernels tested on two consumer GPUs: an NVIDIA GeForce GTX 1660 Ti with 1536 CUDA Cores running at 1500MHz, and an NVIDIA GeForce RTX 3090 with \SI{10496} CUDA Cores running at \SI{1395} MHz.

\begin{table}
\small
  \caption{Time required to calculate the \textit{internal field} on a CPU (single-thread) and two different GPUs launching $F$ independent threads. Units are in seconds.}
  \label{tbl:GPU_internal}
  \begin{tabular*}{\linewidth}{@{\extracolsep{\fill}}llll}
    \hline
    $F_{xy}$(M=1) & Intel I9 & GTX 1660 Ti & RTX 3090\\  
    \hline
    $32\times 32$ & 0.05  & 0.14 &0.26\\
    $64\times 64$ & 0.07 & 0.15 & 0.26\\
    $128\times 128$ & 0.12 & 0.17 &0.26\\
    $256\times 256$ & 0.23 & 0.25 &0.28\\
    \hline \\
    
    \hline
    $F_{xy}$(M=100) & Intel I9 & GTX 1660 Ti & RTX 3090\\ 
    \hline
    $32\times 32$ & 14.5  & 0.28  & 0.39\\
    $64\times 64$ & 58.8 & 0.30 & 0.33\\
    $128\times 128$ & 235 & 0.33 & 0.34\\
    $256\times 256$ & 943 & 0.59 & 0.35\\
    \hline   \\
    
    \hline
    $F_{xy}$(M=1000) &  Intel I9 & GTX 1660 Ti &  RTX 3090\\ 
    \hline
    $32\times 32$ & 2630 (44 m)  & 12.9 & 7.42\\
    $64\times 64$ & 10600 (2.9 h) & 13.5 & 7.26\\
    $128\times 128$ & - - & 13.7 & 7.13\\
    $256\times 256$ & - - & 31.9 & 7.16\\
    \hline \\
    
    \hline
    $F_{xy}$(M=2500) & Intel I9 & GTX 1660 Ti & RTX 3090\\ 
    \hline
    $32\times 32$ & - - &80.9 &48.2 \\
    $64\times 64$ & - - &80.0 &44.9 \\
    $128\times 128$ & - - &90.6 &44.2\\
    $256\times 256$ & - - &206 &46.7 \\
    \hline 
  \end{tabular*}
  \label{tbl:internal_field}
\end{table}

In practice, only the external field needs to be simulated and visualized for field prediction, spectroscopy measurement, etc. Table \ref{tbl:external_field} shows the evaluation results for an external plane.

\begin{table}
\small
  \caption{Time required to calculate the \textit{external field} on a CPU (single-thread) and two different GPUs launching $F$ independent threads. Units are in seconds.}
  \label{tbl:GPU_external}
  \begin{tabular*}{\linewidth}{@{\extracolsep{\fill}}llll}
    \hline
    $F_{xy}$(M=1) & Intel I9 & GTX 1660 Ti & RTX 3090\\ 
    \hline
    $32\times 32$ & 0.05  & 0.15 & 0.27\\
    $64\times 64$ & 0.07 & 0.15 &  0.27\\
    $128\times 128$ & 0.10 & 0.17 &  0.26\\
    $256\times 256$ & 0.16 & 0.24 & 0.24\\
    \hline \\
    
    \hline
    $F_{xy}$(M=100) & Intel I9 & GTX 1660 Ti & RTX 3090\\ 
    \hline
    $32\times 32$ & 0.17  & 0.15  &0.26\\
    $64\times 64$ & 0.44 & 0.15 & 0.26\\
    $128\times 128$ & 1.53 & 0.16& 0.26\\
    $256\times 256$ & 5.87 & 0.23& 0.26 \\
    \hline   \\
    
    \hline
    $F_{xy}$(M=1000) & Intel I9 & GTX 1660 Ti & RTX 3090\\ 
    \hline
    $32\times 32$ & 1.09  & 0.15&  0.24\\
    $64\times 64$ & 3.69 & 0.16& 0.27\\
    $128\times 128$ & 14.5 & 0.17&  0.25\\
    $256\times 256$ & 57.5 & 0.27 & 0.27\\
    \hline \\
        
    \hline
    $F_{xy}$(M=2500) & Intel I9 & GTX 1660 Ti & RTX 3090\\ 
    \hline
    $32\times 32$ &2.46 &0.24 &  0.25\\
    $64\times 64$ &9.09 &0.24 & 0.25\\
    $128\times 128$ &35.6  &0.29 &0.27\\
    $256\times 256$ &142  &0.40  & 0.29 \\
    \hline 
  \end{tabular*}
  \label{tbl:external_field}
\end{table}

\subsection{Optimization of the Eigendecomposition}
By parallelizing the sampling process, the next bottleneck becomes eigendecomposition of $\mathbf{D}$, which is computationally challenging on high-performance systems. The performance of the model scales quadratically with the number of Fourier coefficients $M$ and linearly with the number of the sample layers $L$ (Section \ref{sec:complexity}).

This provides two options for parallelism:
\begin{itemize}
    \item \textbf{Parallelize across $L$}: Since the eigendecompositions are independent across layers, we run all $L$ eigendecompositions in parallel to reduce the complexity to $O(M^3)$.
    \item \textbf{Parallelize Eigendecomposition}: Use a parallel implementation of the eigendecomposition algorithm and run each layer in series.
\end{itemize}
The most efficient option will depend on system architecture, and both can be implemented using a large distributed system. Our profiling of this algorithm (Table \ref{tbl:Intel-MKL}) aims to provide guidance depending on the sample description and system architecture. 

The property matrix $\mathbf{D}^{(\ell)}$ is a complex non-symmetric inverse block diagonal matrix (Equation \ref{eqn:matrix_D}). This limits viable algorithms to implementations of \textit{cgeev} (LAPACK). The open source \textit{Eigen} library provides an appropriate single-threaded eigendecomposition. In addition, both NumPy and Intel MKL have comparable implementations of \textit{geev} using multiple threads simultaneously.

For the simulation, we use the library \textit{Eigen} as the basic data structure to operate blocks in matrices easily. \textit{Eigen} does well in matrix block operations, while the execution time is much longer compared with \textit{Intel-MKL}. As Table \ref{tbl:Intel-MKL}  shows, for a sample with the dimension of $[X, Y, L]=[256, 256, 1]$ simulating on a system with an Intel(R) Core(TM) i9-10900F CPU at a base clock of $2.80$ GHz, we observed that the eigen decomposition using \textit{Intel-MKL} took $99\%+$ less time than it on \textit{Eigen} and obtained subtle advantage over Numpy. We also applied the inversion operation and matrix multiplication in \textit{Intel-MKL} to the model, and the total simulation time decreased by $98.53\%$ and $99.16\%$ respectively compared to the raw Eigen model.

\begin{table}[tbh]
\small
  \caption{Profiling Eigen decomposition for Eigen, Numpy, and Intel-MKL. Units are in seconds.}
  \label{tbl:Intel-MKL}
  \begin{tabular*}{\linewidth}{@{\extracolsep{\fill}}llll}
    \hline
    M & Eigen & Numpy & Intel-MKL\\
    \hline
    100 & 16.4 & 0.11 & 0.10 \\
    400 & 998 & 2.19 & 2.03 \\ 
    900 & - - & 25.1 & 23.2 \\
    1600 & - - & 145 & 143 \\
    2500 & - - & 487 & 461 \\
    \hline 
  \end{tabular*}
\end{table}

\subsection{Simulation accuracy analysis}
\begin{figure}[ht!]\centering
\includegraphics[width=0.75\textwidth]{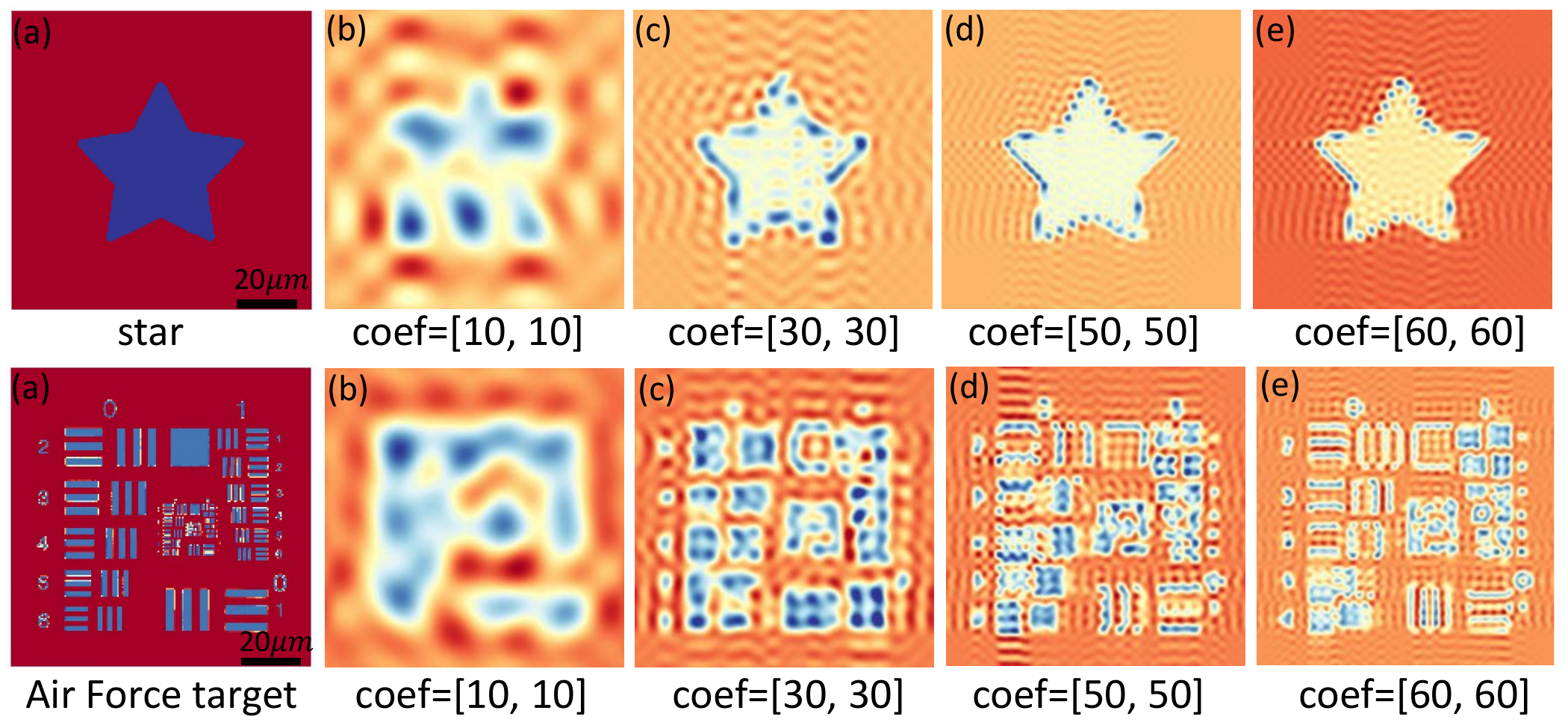}
\caption{The electric field collected from the bottom of the samples for different numbers of Fourier coefficients.}
\label{fig:shapes}
\end{figure}

Since the number of Fourier coefficients $M$ mostly decides the execution time of the model, we show how the $M$ affects the simulation accuracy for different sample shapes. In Fig. \ref{fig:shapes}a, the upper sample "star" and the lower "US Air Force target" have the same resolution (186$\times$186 pixels) and size (\SI{100}{\micro\meter}$\times$\SI{100}{\micro\meter}). With the same incident wavelengths $\lambda=$\SI{390}{\nano\meter}, the Fig. \ref{fig:shapes}b-d show that more complicated sample needs more Fourier coefficients to fully express. 

By adjusting the value of M, we can balance the computational cost of the analysis with the accuracy of the results. As we increase M towards the resolution of the sample, the results become more accurate, but also require more computational resources.

\subsection{Scattering through a cylinder}
\begin{figure}[ht!]\centering
\includegraphics[width=0.75\textwidth]{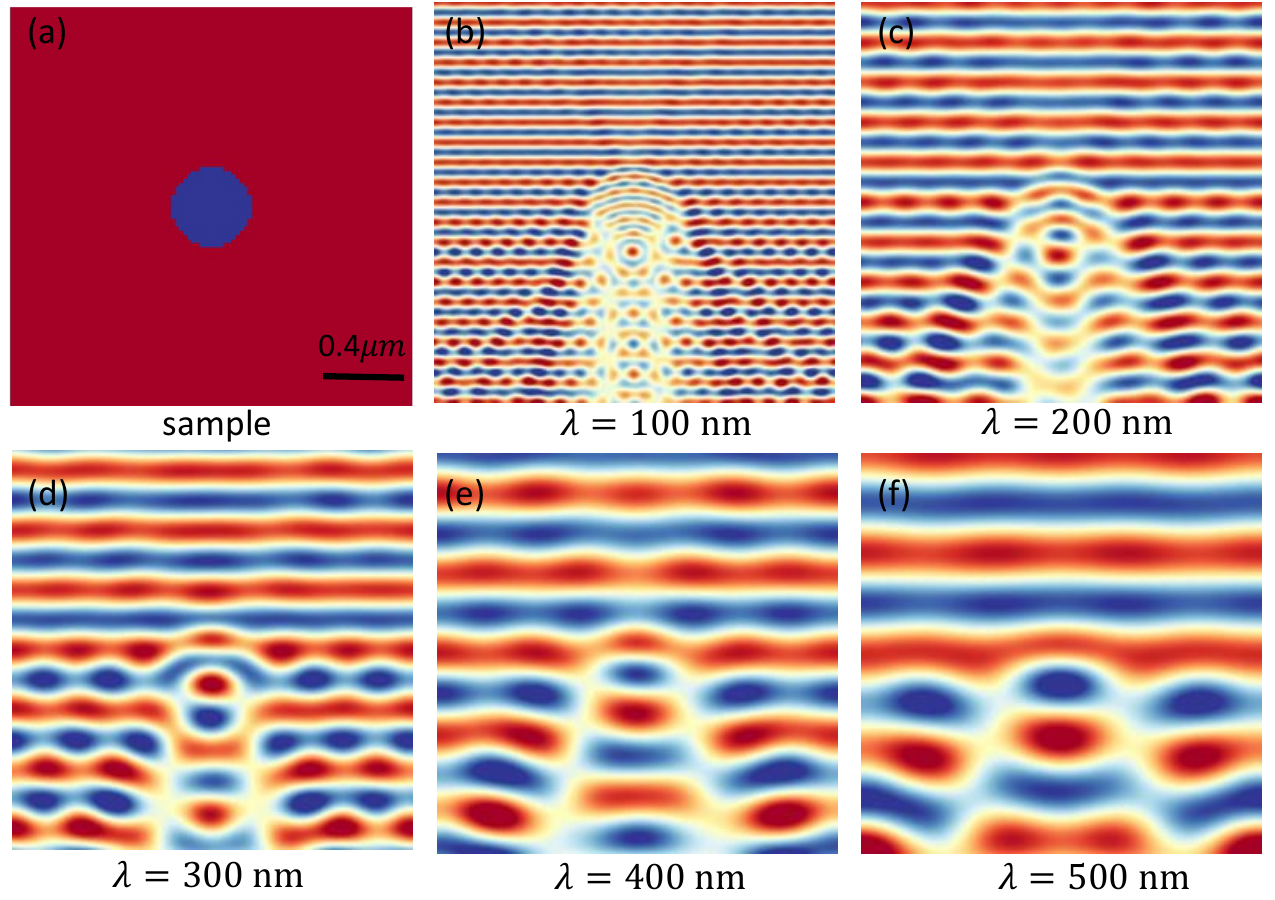}
\caption{(a) shows a $x-z$ cross-section of a cylinder. (b)-(f) are the electric field for the cross-section showing how the scattering happens around a cylinder under the incidence of different wavelengths.}
\label{fig:cylinder}
\end{figure}

The incident plane wave is a $y$-polarized wave with an amplitude of $1$V. The cylinder sample is located at the $x-z$ plane and expands infinitely along the $y$ axis. The original sample in Fig. \ref{fig:cylinder}a is \SI{2}{\micro\meter}$\times$\SI{2}{\micro\meter} with a refractive index of $1$ as background and $1.4+0.05j$ as the circle in the center has a diameter of \SI{400}{\nano\meter}. Fig.\ref{fig:cylinder} b-d shows the electric field $E_y$ around the cylinder under different wavelengths.

\subsection{Absorbance Spectroscopy}
Infrared spectroscopy imaging (IR imaging) is a type of spectroscopic imaging that uses infrared radiation to identify and map the chemical composition of materials. IR radiation is invisible to the human eye, but it can be detected and analyzed using specialized instruments. By shining a beam of IR radiation onto a sample, the IR radiation interacts with the molecules in the sample, and certain wavelengths of IR radiation are absorbed by the molecules. The pattern of absorption is unique to each molecule, so by analyzing the pattern of IR absorption, the type of molecules present in the sample can be identified.

\begin{figure}[ht!]\centering
\includegraphics[width=0.75\textwidth]{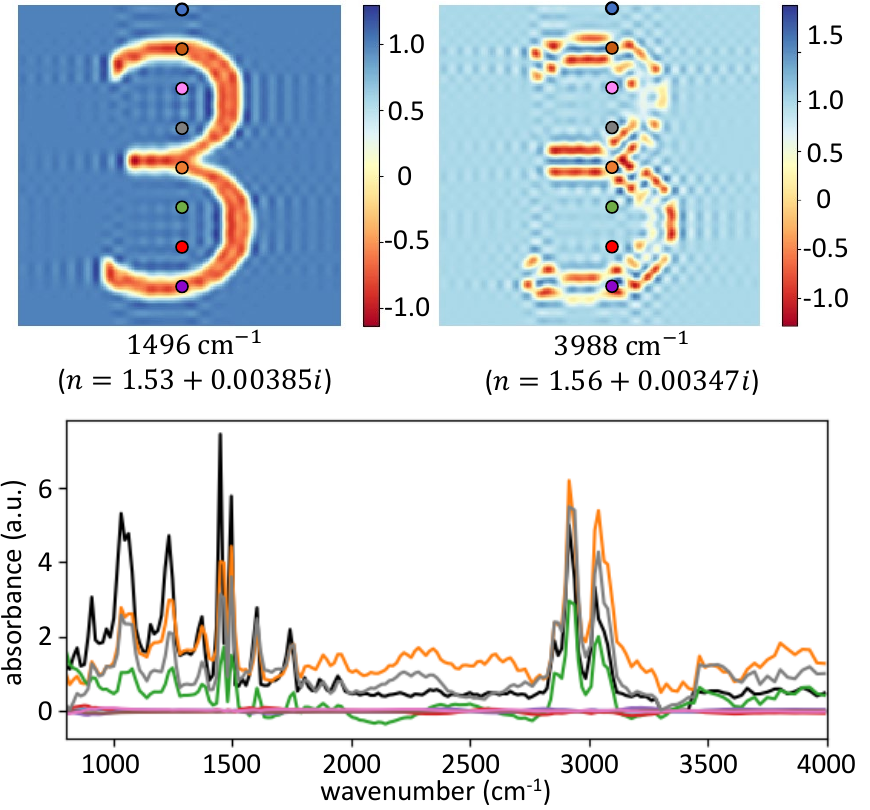}
\caption{Spectroscopy distortions at different wavelengths. }
\label{fig:spectorscopy}
\end{figure}

There are two main types of IR imaging: transmission IR imaging and reflection IR imaging, \textit{transmission IR imaging} and \textit{reflection IR imaging}. In this paper, we detect the intensity of the transmission waves traveling through the sample. Fig. \ref{fig:spectorscopy} shows the spectrum for a character $"3"$ with a resolution of 40$\times$40 pixels at variant positions. Compared to the real spectroscopy curve (black lines in Fig. \ref{fig:spectorscopy}), we can observe different distortions under variant wavelengths caused by scattering.

\subsection{Periodic Heterogeneous three-dimensional sample}
\begin{figure}[ht!]\centering
\includegraphics[width=0.75\textwidth]{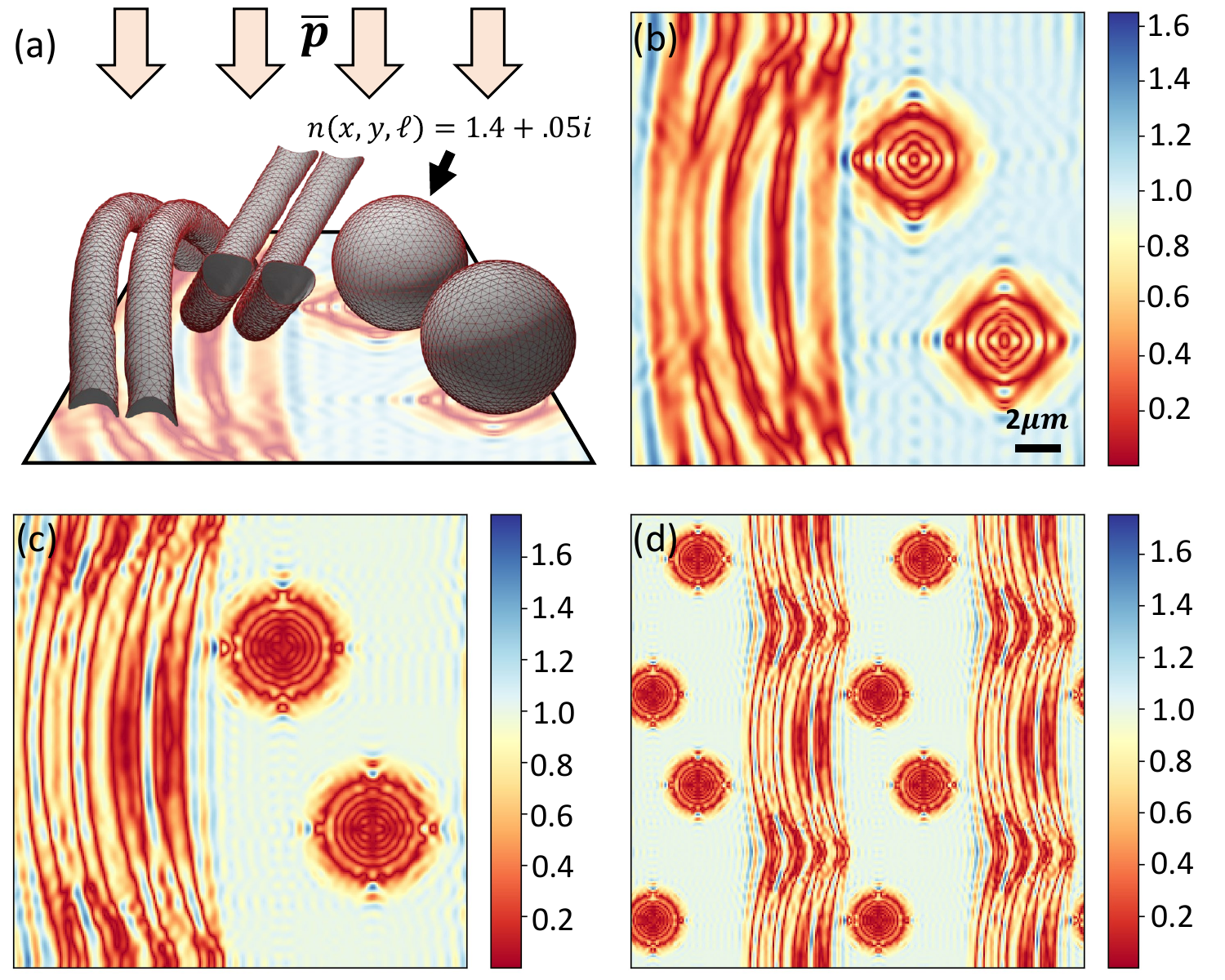}
\caption{(a) shows the sample. (b) is the collected intensity distribution at the lower boundary with the incident wavelength of \SI{625}{\nano\meter}. (c) shows the intensity map at the lower boundary with the incident wavelength of \SI{313}{\nano \meter}. (d) is the periodic display of (c).}
\label{fig:tissue}
\end{figure}

A synthetic tissue sample composed of capillaries and cells is constructed in Fig.\ref{fig:tissue}a, where curve cylinders represent the capillaries and the cells are seen as spheres, where the diameter of the cells is \SI{6.25}{\micro\meter}. With a single wave with a wavelength of \SI{625} {\nano \meter} incident on the sample, Figure \ref{fig:tissue}b shows the intensity collected from the lower $z$ boundary of the sample. Another key advantage of the CW-based model is its periodicity. By efficiently calculating the wave behavior in a single unit cell and then applying it to the entire pattern, the simulation can be just as fast as for a single object. With the incident wavelength of $\lambda=$\SI{313}{\nano \meter}, the pairs c/d in Fig.\ref{fig:tissue} highlight the versatility of our model in handling simulations that require the representation of periodic elements.

\section{Conclusions}
This paper proposes a novel three-dimensional physical model for simulating the electric field distribution around arbitrary sample shapes, utilizing the framework of CW theory. Our key innovation lies in the development of novel connection equations, which dramatically reduce the dimensionality of the resulting linear system (Fig. \ref{fig:linear_functions_Hete}) by a factor of $L$, enabling efficient simulation of multi-layered samples. 

Furthermore, we optimize the simulation process by leveraging the high-performance computing capabilities of \textit{Intel-MKL} (Fig.\ref{tbl:Intel-MKL}). Compared to the standard Eigen library, Intel MKL's multi-threaded Eigensolver compresses the simulation time to a practical 15 minutes for commonly encountered scenarios while maintaining acceptable accuracy. To expedite the visualization stage, we implement CUDA computing, a parallel programming paradigm that utilizes multiple computational cores simultaneously. This technique enables real-time visualization of dense volumetric data ($256\times256\times 256$ pixels) within seconds to minutes, depending on the number of employed Fourier coefficients.

The combined strength of efficient field simulation and rapid visualization empowers our model to effectively analyze light scattering phenomena within realistic timeframes, overcoming convergence issues typically encountered in complex scenarios. This paves the way for diverse applications, including the visualization of physical phenomena, the design of novel optical systems, and the validation of scattering results obtained through current microscopy techniques.

\section*{Author Contributions}
RS developed the connection equation additions to the CW-based model and optimized the software. RR designed the mathematical model. DM advised on software development and optimization. All three authors contributed to writing the manuscript. 

\section*{Conflicts of interest}
DM is a stakeholder in SwiftFront, LLC.

\section*{Acknowledgements}
This work was funded in part by the National Institutes of Health / National Heart, Lung, and Blood Institute (NHLBI) \#R01HL146745 and the National Science Foundation CAREER Award \#1943455, the NLM Training Program in Biomedical Informatics and Data Science T15LM007093 (RR), and the Cancer Prevention and Research Institute of Texas (CPRIT) \#RR170075 (RR).
\bibliography{sample}





\end{document}